\documentclass[11pt,twoside]{article}
\usepackage{asp2014}

\aspSuppressVolSlug
\resetcounters
\begin{document}

\title{A Novel JupyterLab User Experience for Interactive Data Visualization\footnote{}}

\author{Peter K.\ G.\ Williams$^{1,2}$}
\author{Jonathan Carifio$^1$}
\author{Henrik Norman$^3$}
\author{A.\ David Weigel$^4$}
\affil{$^1$Center for Astrophysics | Harvard \& Smithsonian, Cambridge, MA, USA; \email{pwilliams@cfa.harvard.edu}}
\affil{$^2$American Astronomical Society, Washington, DC, USA}
\affil{$^3$Winter Way AB, Lerum, V\"asterg\"otland, Sweden}
\affil{$^4$US Space \& Rocket Center, Huntsville, AL, USA}

\paperauthor{Peter~K.~G.~Williams}{pwilliams@cfa.harvard.edu}{0000-0003-3734-3587}{Center for Astrophysics | Harvard \& Smithsonian}{}{Cambridge}{MA}{02138}{USA}
\paperauthor{Jonathan~Carifio}{jonathan.carifio@cfa.harvard.edu}{0000-0002-7759-2601}{Center for Astrophysics | Harvard \& Smithsonian}{}{Cambridge}{MA}{02138}{USA}
\paperauthor{Henrik~Norman}{henrik@winterway.eu}{0000-0003-4189-3450}{Winter Way AB}{}{Lerum}{V\"asterg\"otland}{443~38}{Sweden}
\paperauthor{A.~David~Weigel}{davidw@spacecamp.com}{0000-0002-8026-2291}{US Space \& Rocket Center}{}{Huntsville}{AL}{35805}{USA}




\begin{abstract}
In the Jupyter ecosystem, data visualization is usually done with ``widgets''
created as notebook cell outputs. While this mechanism works well in some
circumstances, it is not well-suited to presenting interfaces that are
long-lived, interactive, and visually rich. Unlike the traditional Jupyter
notebook system, the newer JupyterLab application provides a sophisticated
extension infrastructure that raises new design possibilities. Here we present a
novel user experience (UX) for interactive data visualization in JupyterLab that
is based on an ``app'' that runs \textit{alongside} the user's notebooks, rather
than widgets that are bound \textit{inside} them. We have implemented this UX
for the AAS WorldWide Telescope (WWT) visualization tool. JupyterLab's messaging
APIs allow the app to smoothly exchange data with multiple computational
kernels, allowing users to accomplish tasks that are not possible using the
widget framework. A new Jupyter server extension allows the frontend to
request data from kernels asynchronously over HTTP, enabling
interactive exploration of gigapixel-scale imagery in WWT. While we have
developed this UX for WWT, the overall design and the server extension are
portable to other applications and have the potential to unlock a variety of new
user activities that aren't currently possible in ``science platform''
interfaces.\footnote[1]{%
This submission also includes material corresponding to tutorial T04,
``Interactive Visualization in the Age of the Science Platform: Huge FITS
Images in JupyterLab with AAS WorldWide Telescope''.}
\end{abstract}

\section{Introduction}

Many astronomical organizations are building, or have already built, ``science
platforms'': online environments exposing both data and computation to allow
researchers to make use of major data sets from afar. Examples include SciServer
\citep{the.sciserver}, NOIRLab's Astro Data Lab \citep{the.datalab}, ESA
Datalabs \citep{the.esa.datalabs}, and the Rubin Science Platform
\citep{the.rsp}. The motivation for the creation of these platforms is
well-understood: modern astronomical datasets are so big that it is impractical
for researchers to simply ``download the data'' and analyze them. Instead, the
researchers must bring their analysis \textit{to} the data.

Paired with this motivation are the factors that make the creation of science
platforms possible: perhaps most importantly, the immense sophistication of the
modern web development platform. Current web browsers are sophisticated tools
for building interactive, multimedia applications, and there exists a rich
ecosystem of software tools for implementing web applications. Among astronomers
and other scientists, the Jupyter system is particularly important. While one
important aspect of the now-established Jupyter notebook system
\citep{the.jupyter} is the design of the ``notebook'' user experience (UX)
paradigm, just as important is the fact that Jupyter was built as a web
application, targeting HTML and JavaScript rather than Windows or macOS.

The move to web-based scientific data analysis presents both opportunities and
challenges. Regarding data visualization, especially interactive data
visualization, one major challenge is simply achieving feature parity with the
status quo. While some kinds of software are easily translated from existing
platforms to the web platform, that is \textit{not} the case for visualization
applications, which are generally interactive and have complex I/O patterns. The
associated opportunity is that newer web-based UX paradigms and workflows have
the potential to be superior to existing ones. While it remains to be seen to
what extent this potential can be realized in practice, the network-native
nature of the web enables workflows that achieve unprecedented levels of
openness, collaborativity, and decentralization.

\section{Jupyter and Widgets}

Since its first public release, the Jupyter environment has provided a UX
paradigm for web-based data visualization: the notebook widget, as implemented
in the \textsf{ipywidgets}
package\footnote{\url{https://ipywidgets.readthedocs.io/}}. The mental model
underlying widgets is intuitive: while the output of an executed notebook cell
is generally textual, it can actually consist of a variety of forms, such as
images, audio, or arbitrary embedded HTML. Interactive Jupyter widgets use
Jupyter's ``comms'' system to communicate changes between the underlying
computational kernel (often, but not necessarily, running Python) and the user's
web browser (implementing the widget display in JavaScript). The widget system
is based on a model/view/controller (MVC) framework in which there can be
multiple HTML views of the same JavaScript model, and the latter is what is
synchronized with the kernel.

Although Jupyter widgets can be used to achieve complex effects, there are some
limitations inherent to the system's design. A somewhat banal but salient issue
is that since widgets are tied to cells, as one works in a notebook, they tend
to scroll off the screen, requiring scrolling back and forth to keep using them.
There are workarounds to avoid this issue such as
\textsf{jupyterlab-sidecar}\footnote{\url{https://github.com/jupyter-widgets/jupyterlab-sidecar}},
but these require the user to take additional steps to activate them. In the
case of \textsf{jupyterlab-sidecar} this includes modifications to notebook code
that have to be duplicated or removed depending on whether or not the sidecar
can and should be used.

Furthermore, widgets must be launched in code, by a running kernel. In many
cases this is not a significant limitation, but it does preclude or hamper
certain workflows. For instance, in a science platform environment, one might
wish to simply select a dataset in a file browser and open it for visualization.
To accomplish this using a widget requires a relatively involved set of steps:
in general one will have to create a notebook, load the appropriate Python
modules, copy the file's name into the session, and finally run the necessary
code to display it.

Finally, widgets are tied to a single running kernel through Jupyter's comms
system. Although widgets can have their state serialized and restored through
kernel restarts, this is only practical for small state sizes, not (e.g.) large
images loaded for interactive visualization. Because widgets are tied to
individual kernels, data from multiple notebooks cannot be combined into the
same visualization, and in the JupyterLab environment (see next section),
application components besides notebooks cannot interact with widgets.

\section{The JupyterLab Computational Environment}

The release of the JupyterLab software stack --- to be distinguished from the
``classical'' Jupyter notebook --- marked a major expansion in possibilities for
interactive data visualization on the web. While the project's branding may not
distinguish it substantially from previous Jupyter products, the JupyterLab
system represents a much more ambitious and sophisticated vision than what came
before.

\begin{figure}[tb]
  \includegraphics[width=\linewidth]{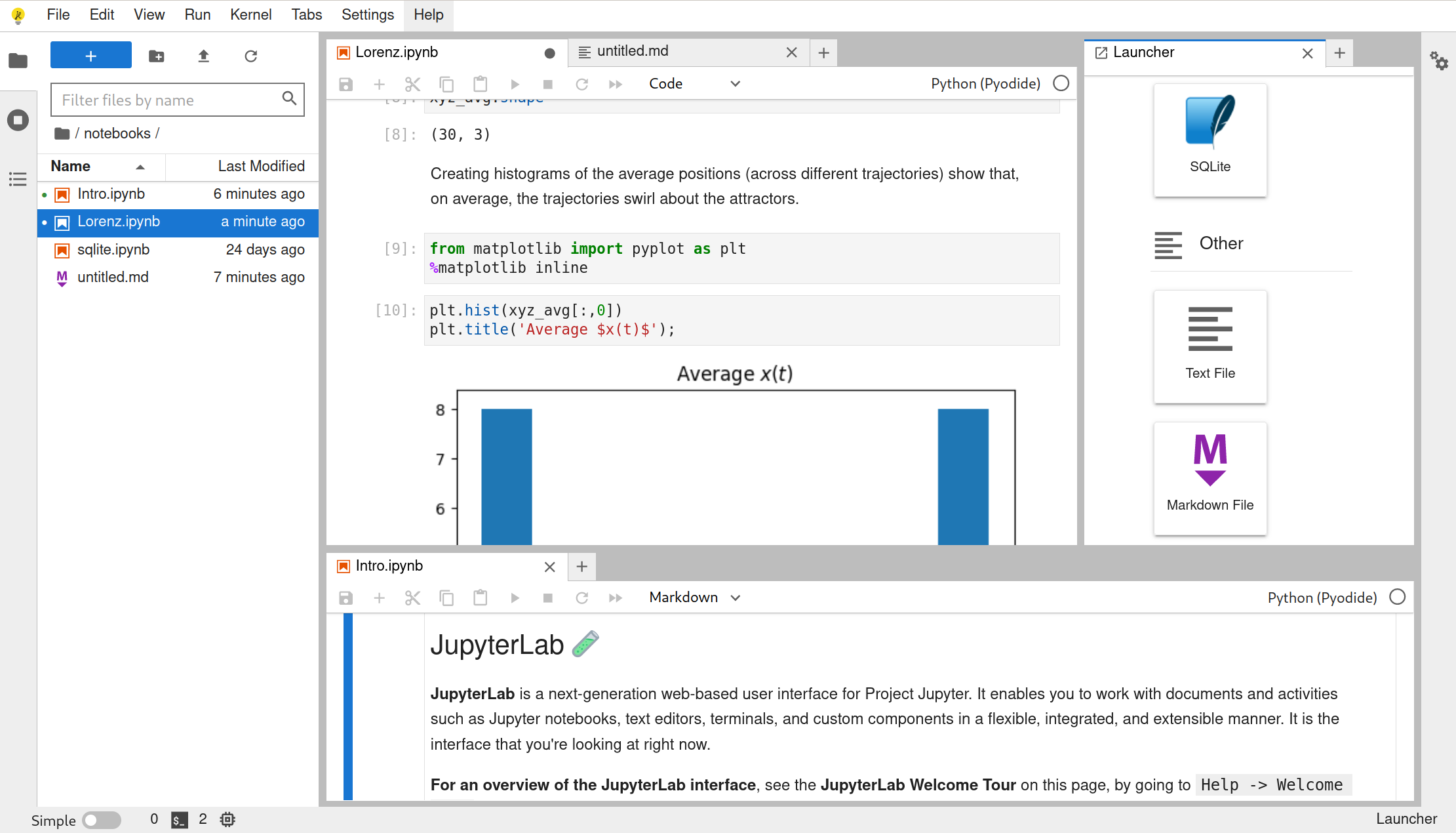}
  \caption{Screenshot of a representative JupyterLab environment. Unlike the
classical Jupyter web UI, multiple notebooks may be opened and arranged in the
main view area. The sidebar and menu bars provide access to a variety of
associated features not available in the classical app. Notebook widgets are
supported with the help of a standard extension.}
  \label{f.jlab}
\end{figure}

The web frontend of the classical Jupyter system was designed to run notebooks,
although it did gain additional related functionality over time. In contrast,
JupyterLab is designed to be a modular, extensible, web-based environment for
computation. While it supports notebooks as a part of that functionality, the
complete system is much more flexible. Superficially, this can be seen in the
JupyterLab UI (\autoref{f.jlab}), which adds a tabs-and-frames system for
managing different tools such as notebooks, terminals, and code editors. Under
the hood, nearly every component of JupyterLab is built using its extension
mechanism, which can be used to add new launcher buttons, menu commands, frame
types, and more.

While JupyterLab supports classical Jupyter widgets with the assistance of the
\textsf{jupyterlab-manager} extension provided by \textsf{ipywidgets}, the
extensibility of the JupyterLab environment makes it possible to implement new
visualization UX paradigms that go beyond the widget paradigm.

\section{The JupyterLab ``App'' User Experience Paradigm}

\begin{figure}[tb]
  \includegraphics[width=\linewidth]{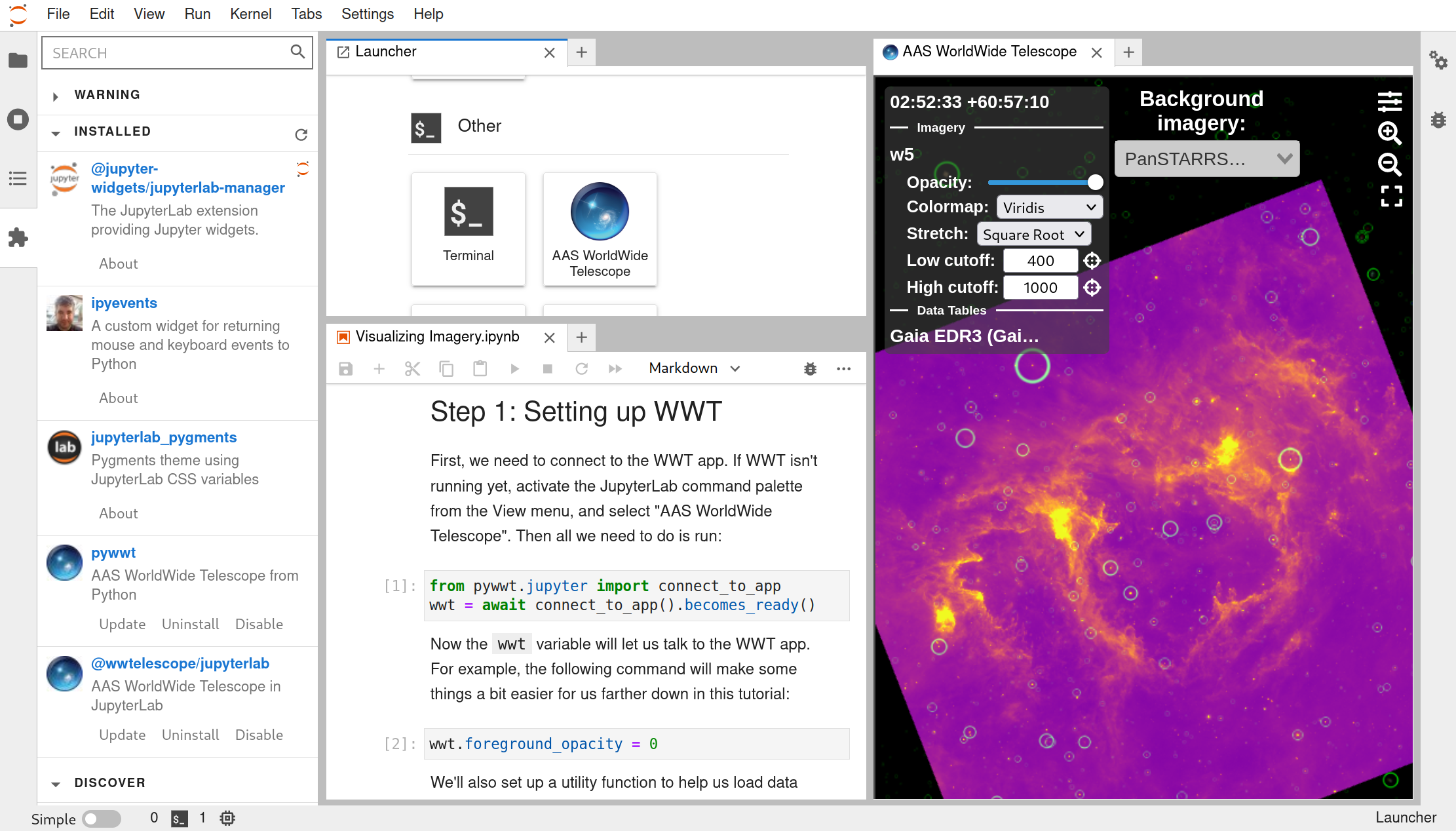}
  \caption{Screenshot of the WWT JupyterLab app. The main WWT interface is
contained within its own tab frame, and so can be repositioned and resized by
the user. A button in the ``Launcher'' view (top-middle) makes opening WWT a
one-click operation. Connecting a Python kernel to the app requires two lines of
code (bottom-middle). However, because the app itself has a well-developed user
interface (right), many actions can be performed without an active kernel
connection.}
  \label{f.app}
\end{figure}

AAS WorldWide Telescope (WWT; \citealp{the.wwt}) is an application for
interactive astronomical data visualization that specializes in the combination
of multiple datasets on the sky, including both images and catalogs of arbitrary
size. While WWT has provided a Jupyter widget since 2017 via the \textsf{pywwt}
package\footnote{\url{https://pywwt.readthedocs.io/}}, the constraints of the
widget UX have not been a good match for the style of use that WWT encourages.
Motivated by this mismatch and the new capabilities of JupyterLab, the WWT team
has implemented a new UX design in the extension
\textsf{@wwtelescope/jupyterlab}\footnote{\url{https://github.com/WorldWideTelescope/wwt-jupyterlab}}.
This implementation is an instance of a new JupyterLab UX design pattern that we
refer to as the ``app'' model (\autoref{f.app}). It can be tried online using
the MyBinder service\footnote{\url{https://bit.ly/pywwt-notebooks}}
\citep{the.binder}.

The hallmark of the ``app'' paradigm is that the visualization application is
chiefly implemented inside its own JupyterLab extension, rather than through the
\textsf{ipywidgets} framework. It is straightforward to extend the JupyterLab
shell so that such an app can easily by opened through a button in the
JupyterLab launcher, the command bar, or by selecting appropriate files in the
file explorer. This makes it possible to launch the app without requiring a
kernel, and also allows JupyterLab's tab manager to be used to set up custom
views, such as placing the app and a notebook in a side-by-side configuration.

Rather than launching the app \textit{from} a notebook kernel, instead kernels
are viewed as connecting \textit{to} the app. Multiple notebooks can connect to
the same app, allowing different data sources to be combined. JupyterLab's
communications framework allows bidirectional messaging between an app and
kernels, allowing data to flow both from a kernel to the app (e.g., ``display
this small catalog'') and the other way around (e.g., ``retrieve the coordinates
of the current view center'').

Because of this bidirectional communication, apps may rely on notebooks to
provide computational support to accomplish effects that would be difficult to
achieve in a freestanding web application. For instance, in the case of WWT,
there are many opportunities for data discovery and retrieval through Virtual
Observatory (VO; \citealp{the.vo}) protocols. However, to the best of our
knowledge there are limited, if any, VO protocol implementations available in
JavaScript, so that it would be expensive to provide VO support inside the WWT
application per se. Meanwhile there are many VO implementations available in
Python, such as \textsf{pyvo}\footnote{\url{https://pyvo.readthedocs.io/}}. A
pragmatic approach would be to allow the app to use a connected kernel to make
VO requests and parse the responses, minimizing the amount of new logic required
in the web app. This approach can address some, but not all, of the costs
associated with porting existing visualization apps to the web platform.

Not explored in the current WWT implementation is the possibility of an app
providing even stronger notebook integration by generating code for the user to
run. For instance, on startup the app could offer to copy/paste the generic two
lines of code needed to connect a kernel to itself, or it could emit code that
recreated the current viewer location to create notebooks that reliably
reproduce the results of an interactive session. Apps offering such features
should be designed to recognize the fact that not all Jupyter kernels are
Python-based.

Communications between apps and kernels use Jupyter's messaging system. In the
generic ``science platform'' case, this involves relaying JSON messages from a
kernel, likely running in a remote data center, to the user's browser, via the
Jupyter web server, which may run at some intermediate location. While Jupyter's
infrastructure masks much of the complexity underlying this setup, messages do
arrive asynchronously and unreliably, necessitating careful protocol design.
Message-based control systems are well-suited to this environment and are
well-established in astronomy with protocols such as SAMP \citep{the.samp}.

In the particular case of WWT, the app is a singleton: only one WWT frame can be
open in a JupyterLab session at once. This is a UX design choice; there is no
technical limitation preventing multiple frames of the same app from being
opened.

\section{Visualizing Big Data in the App Paradigm}

The app paradigm is well-suited to applications that involve exploratory
visualization of large datasets. In the case of WWT, an application of
particular interest is the display of large (gigapixel-scale) images.

Web-based, interactive visualization of such datasets requires that they be
processed into a format that allows the client to display something without
needing to download the entire dataset in advance. In WWT, images that span a
substantial solid angle must be converted into the TOAST \citep{the.toast} or
HiPS \citep{the.hips} formats, both of which involve breaking the image data
into tiles that are then hierarchically downsampled to lower resolutions. Images
that contain a large number of pixels but do not span a large area can be
represented using a gnomonic (WCS \texttt{TAN}) projection with a similar
hierarchical tiling scheme. The software package
\textsf{toasty}\footnote{\url{https://toasty.readthedocs.io/}} can generate
these formats from typical astronomical datasets. In other application domains,
the precise data formats may vary but processes analogous to the tiling and
downsampling steps will inevitably apply.

For large, public datasets, it makes sense to precompute these
visualization-friendly outputs and serve them from a well-known location. For
browser-based display the natural service mechanism is HTTP, and indeed the WWT
app has a sophisticated HTTP-based I/O subsystem for retrieving tile data
on-the-fly as the user navigates an image. (Here and elsewhere, ``HTTP'' should
be read to mean ``HTTP or HTTPS''.)

\begin{figure}[tb]
  \includegraphics[width=\linewidth]{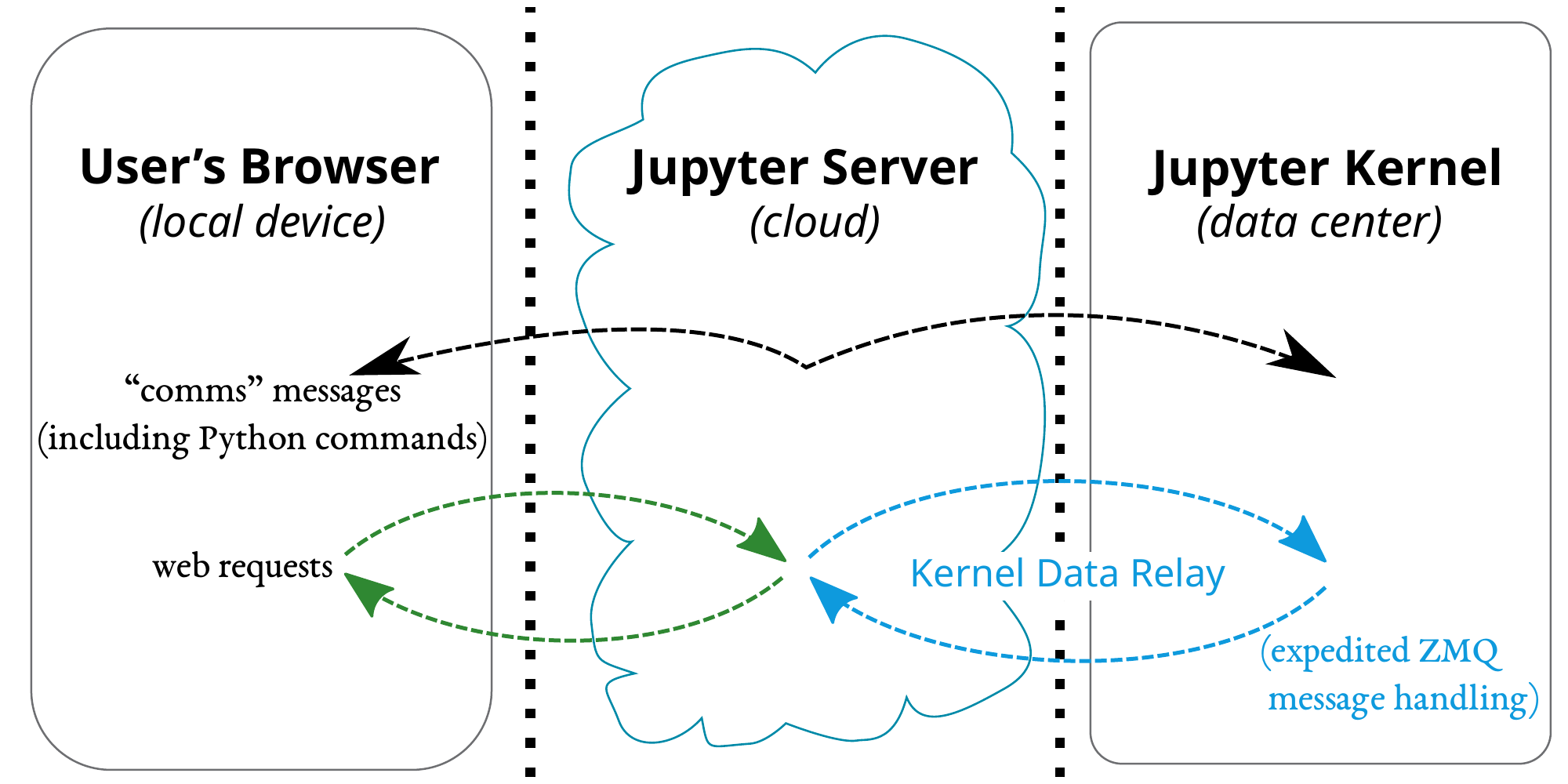}
  \caption{Schematic of messaging in the Jupyter architecture. The Jupyter(Lab)
frontend runs in the user's browser and communicates with the server over HTTP.
The server communicates with the kernels (only one shown here for clarity),
which may be running on a separate network, using ZeroMQ. The frontend can
communicate with kernels indirectly using ``comms'', which tunnel ZeroMQ
messages to the browser using WebSockets. The WWT Kernel Data Relay adds a
mechanism to, in effect, tunnel HTTP requests to one or more kernels using
ZeroMQ.}
  \label{f.comms}
\end{figure}

In the ``science platform'' use case, it is also likely that users will obtain
or generate new datasets that they wish to visualize in the app. In the Jupyter
architecture these datasets will be ``kernel-backed'': the kernel process is the
one that has direct access to the data. These data need to be made available to
the app frontend. Presuming that the frontend has an HTTP-based I/O subsystem,
it would be desirable to make such data available over HTTP --- but this is not
actually possible in the stock Jupyter architecture. Frontend code can only
communicate with the Jupyter \textit{server} using HTTP, and the server and
kernels may be running in different environments (\autoref{f.comms}).

While it may be possible to avoid this problem by implementing a comms-based I/O
subsystem, this is not a feasible solution for WWT. We therefore implemented a
small Jupyter server extension,
\textsf{wwt\_kernel\_data\_relay}\footnote{\url{https://wwt-kernel-data-relay.readthedocs.io/}}
(KDR), that allows kernels to register themselves as handling certain HTTP
requests and provides server hooks to relay those requests and their responses
over the Jupyter ZeroMQ messaging system. The protocol is designed to be as
minimal as possible and should be applicable to other scenarios where a Jupyter
frontend needs to obtain data from a kernel over HTTP.

A further difficulty is that the default
\textsf{ipykernel}\footnote{\url{https://github.com/ipython/ipykernel}} Python
kernel implementation only processes most ZeroMQ messages while it is
executing a code cell. If an app like WWT is being used to explore a large
kernel-backed dataset, this means that data requests will pile up until the user
happens to run code in the right kernel, leading to an unacceptable UX. The
\textsf{pywwt} implementation of the KDR protocol patches the \textsf{ipykernel}
message handler to allow ``expedited'' handling of specially-marked messages.
The KDR server extension uses such markers to avoid this problem. An un-patched
message handler will simply be oblivious to the requests for expedited handling.

\section{Discussion}

It would be inappropriate to describe the app UX paradigm as flatly ``better''
or ``worse'' than the widget paradigm. Each approach has strengths and
weaknesses for various use cases. For long-lived, interactive applications, we
believe that the app design has many advantages, including easy launch, a
reliable and intuitive way to set up a side-by-side view of code and data, and
the ability to combine data from multiple kernels. On the other hand, unlike a
widget, an app cannot be serialized into a saved notebook, and its UI is not
visually connected with the specific lines of code that created it. Finally,
widgets are probably somewhat easier to implement, although we hope that the
open-source nature of the WWT JupyterLab app and new tools such as the KDR make
it easier for future implementers to create apps based on WWT's example.

\acknowledgments

This material is based upon work supported by the National Science Foundation
under Grant No.\ OAC-2004840. Work on the AAS WorldWide Telescope project has
been funded by the American Astronomical Society, the NSF, Space Telescope
Science Institute, and other supporters.

\bibliography{C02}

\end{document}